\documentstyle[11pt, aaspp4, psfig, tighten]{article}

\def\sp{\hspace{1.5pt}}
\def\etal{{\it et~al.}}
\def\amin{\ifmmode^{\prime}\else$^{\prime}$\fi}
\def\asec{\ifmmode^{\prime\prime}\else$^{\prime\prime}$\fi}

\def\simgt{\lower.5ex\hbox{$\; \buildrel > \over \sim \;$}}
\def\simlt{\lower.5ex\hbox{$\; \buildrel < \over \sim \;$}}

\newcommand\ns{neutron star}
\newcommand\NS{NS}

\newcommand\rcw{RCW 103}
\newcommand\src{\hbox{1E\hspace{1.5pt}161348$-$5055}}
\newcommand\geleven{\hbox{G\hspace{1.5pt}11.2$-$0.3}}

\newcommand\einstein{{\it Einstein}}
\newcommand\ASCA{{\it ASCA}}
\newcommand\asca{{\it ASCA}}
\newcommand\ROSAT{{\it ROSAT}}
\newcommand\rosat{{\it ROSAT}}
  
\received{6 Jun 1997}

\lefthead{Gotthelf, Petre \& Hwang}
\righthead{Supernova Remnant RCW 103}

\begin{document}

\title{\Large\bf The nature of the radio-quiet compact \\ X-ray source in \hbox{SNR RCW 103}}

\author{E. V. Gotthelf$^{1,2}$, R. Petre$^{1}$ \&\ U. Hwang$^{1,3}$}

\affil{$^1$ NASA/Goddard Space Flight Center, Code 660.2, Greenbelt MD, 20771, USA}
\altaffiltext{2}{Universities Space Research Association}
\affil{$^3$ University of Maryland, College Park, MD, USA}

\begin{abstract}

We consider the nature of the elusive \ns\ candidate \src\ using X-ray
observations obtained with the \asca\ Observatory. The compact X-ray
source is centered on the shell-type Galactic supernova remnant \rcw\
and has been interpreted as a cooling \ns\ associated with the
remnant. The X-ray spectrum of the remnant shell can be characterized
by a non-equilibrium ionization (NEI) thermal model for a shocked
plasma of temperature $\rm{kT} \sim 0.3$ keV. The spectrum falls off
rapidly above 3 keV to reveal a point source in the
spectrally-resolved images, at the location of \src. A black-body
model fit to the source spectrum yields a temperature $\rm{kT} = 0.6$
keV, with an unabsorbed 0.5--10 keV luminosity of L$_{X} \sim 10^{34}$
erg s$^{-1}$ (for an assumed distance of $3.3$~kpc), both of which are
at least a factor of $2$ higher than predicted by cooling \ns\ models.
Alternatively, a power-law model for the source continuum gives a
steep photon index of $\alpha \sim 3.2$, similar to that of other
radio-quiet, hard X-ray point sources associated with supernova
remnants. \src\ may be prototypical of a growing class of radio-quiet
\ns s revealed by \asca; we suggest that these objects account for
previously hidden \ns s associated with supernova remnants.

\end{abstract}

\keywords{stars: individual (RCW 103, \src) --- stars: neutron --- 
supernova remnants --- X-rays: stars}

\section{Introduction}

Neutron stars (NS) are thought to be born as the stellar remnants of
supernova explosions involving massive star progenitors.  It has long
been assumed that NSs begin life as fast $\sim 10$ ms radio pulsars
emitting non-thermal radiation, with the Crab and Vela pulsars as the 
canonical examples. Thus it is remarkable that few ($\sim 10$) of
the $\sim 200$ known Galactic SNR have pulsars associated with them
(see Kaspi 1997), even though a large fraction of supernova explosions
should have produced \NS s.  Perhaps we are seeing only those young SNR
associated pulsars beamed in our direction or those with sufficiently
large radio luminosities to have been detected (Lorimer 1993); or that
pulsars are born with large ``kick'' velocities, launching them far
from their remnants (Lyne \& Lorimer 1994). But evidence of
radio-loud ``plerions'' associated with young pulsars suggest they
should be easily detectable near identified SNRs (Weiler \& Sramek
1988).  Here we present a radio-quiet \NS\ candidate of moderate X-ray
luminosity associated with the shell-type SNR \rcw. Objects of this
type may or may not share a common origin, but their lack of radio
emission suggests that a reconsideration of pulsar evolution is
needed; they might partially account for the missing young \NS s
associated with SNRs.

The X-ray source \src\ was first detected with the high resolution
imager (HRI) on-board the \einstein\ Observatory as a faint,
unresolved source located near the center of \rcw\ (Tuohy \& Garmire
1980). Although the location of \src\ suggests an association with
\rcw, systematic optical and radio searches over the years were unable
to find a counterpart (Tuohy \& Garmire 1980, Tuohy \etal\ 1983,
Dickel \etal\ 1996). A recent radio pulsar search using the Parkes
radio telescope also produced a null result (Kaspi \etal\ 1996).  With
the lack of observational evidence for an accreting binary, Tuohy
\etal\ (1983) initially proposed \src\ as an isolated \NS\ emitting
thermal radiation, the first to be discovered.  Subsequent
observations using the imaging proportional counter (IPC) on-board
\einstein\ failed to confirm its existence, ostensibly due to the
IPC's poor spatial resolution (Tuohy \& Garmire 1980); and
non-detections by the instruments on-board the \ROSAT\ Observatory
over the intervening years are attributed to degraded sensitivity of
the off-axis target position (Becker \etal\ 1993).  Until the present
study, no systematic broad-band X-ray spectral analysis had been
possible.

\rcw\ (G332.4-0.4) is an X-ray bright, nearly circular ($\sim 9'$
diameter), shell-type Galactic SNR.  Its distance is
uncertain. Caswell \etal\ (1975) derived a value of 3.3 kpc from 21 cm
Hydrogen line absorption measurements; an optical extinction study by
Leibowitz \& Danziger (1983) in the vicinity of \rcw\ yields a value
of 6.6 kpc. Herein we adopt the former, more generally accepted,
value, for consistency with previous studies. \rcw\ is similar to
other young remnants in having a bright radio shell traced by the
optical filaments (van den Bergh 1978; Dickel \etal\ 1996). Its small
inferred diameter of ($\sim 10$ pc) and the high velocity dispersion
of its optical filaments ($\sim 900$ km s$^{-1}$; Tuohy, \etal, 1979)
are also consistent with a young ($\sim 1000$ yrs) remnant (Nugent
\etal\ 1984).

In this Letter we present confirmation of the existence of \src,
characterize its spectrum, and reconcile the earlier non-detections
with the \asca\ data. Our detection is confirmed by a new {\it
on-axis} \ROSAT\ HRI observation recently made available in the public
archive. In addition, we use spectral evidence from the nebula to
portray a consistent picture of \src\ as a young, Type II SNR with an
embedded \NS. We present \src\ as proto-typical of a growing class of
remnants with central radio-quiet \NS s, which have hard spectra, but
lack evidence for plerionic emission from a synchrotron
nebula. Members of this class may also include CTB\sp109 (Corbet
\etal, 1995), Puppis-A (Petre \etal, 1996), G296.5+10.0 (Mereghetti
\etal\ 1996) and Kes 73 (Gotthelf \& Vasisht 1997).

\section{Observations}

A day-long observation of \rcw\ was performed using the \ASCA\
Observatory (Tanaka \etal\ 1994) on Aug 17, 1993, during the
Performance Verification (PV) phase of the mission. We used archival
data from both pairs of on-board instruments, the solid-state
spectrometers (SIS) and the gas scintillation spectrometers (GIS). The
SIS detectors offer $\sim 2 \%$ spectral resolution at 6 keV
(resolution $\sim \rm{E}^{-1/2}$ over its $0.4-12$ keV bandpass).  The
spatial resolution is limited by the mirror point spread function
(PSF), whose sharp core gives a FWHM of $\sim 1'$ but large wings
produce a half power diameter of $\sim 3'$ (Jalota \etal\ 1993).
These wings are problematic when extracting and modeling the spectrum
of a point source embedded in diffuse emission. All SIS data were
acquired in 4 CCD BRIGHT mode and screened using the standard REV1
processing.

We also extracted GIS data to search for pulsations and to help
constrain the high energy spectrum. The GIS profits from more screened
observing time and greater sensitivity above 2 keV than the SIS, but
with poorer spectral ($\sim 8\%$ at 6 keV) and spatial (FWHM $\sim
3^{\prime}$) resolution. The GIS data were collected in the
highest time resolution mode ($0.5 \ \rm{ms}$ or $64 \ \mu\rm{s}$,
depending on data acquisition mode), with reduced spatial ($1' \times
1'$ pixels) and spectral bining ($\sim 12$ eV per PHA channel). A total
of 47 ks of screened SIS data and 71 ks of
screened GIS data were acquired from the observation.

Each SIS consists of four X-ray CCDs tiled in a square with a gap
between each pair of chips. The $22^{\prime}\times22^{\prime}$ fields
of view of the two instruments are slightly offset ($\sim 1^{\prime}$)
in the direction of the larger of the two gaps.  During this early PV
observation, \src\ was placed at an unfortunate location on the SIS,
straddling the inter-CCD gaps.  To compensate for aspect jitter near
the gaps, we applied exposure corrections in constructing images.
Figure 1 presents the unambiguous \ASCA\ detection of \src, the first
since its discovery. The hard-band ($> 3.0$ keV) image (Fig. 1a)
shows a distinct feature centered on the remnant, whose spatial
distribution is consistent with that from a point source. In stark
contrast, the soft-band ($0.5-1.5$ keV) image shown in Fig. 1b is
dominated by diffuse emission from \rcw, and reproduces low-resolution
X-ray images of \rcw\ acquired with previous missions. Both images are
centered on the bright peak in the hard-band map. \src\ is also
evident in the hand-band GIS image, albeit at the poorer spatial GIS
resolution.

To locate \src\ with improved accuracy we use the arcsec astrometry
of the radio maps (Dickel \etal\ 1996) to register the SIS image of
the remnant, which is found to correlates well with the radio
morphology. The derived J2000 coordinates are R.A. $ 16^h 17^m
35^s.5$, Dec. $-51{^\circ} 02^{\prime} 21$ degrees, $7^{\prime\prime}$
away from the \einstein\ location, and within the expected offset of
the combined \einstein\ error circle and \asca\ measurement accuracy
(see Gotthelf 1996). The centroid of the source is aligned with the
radio depression and appears to be within an arc-minute of the
inferred radio center of the remnant.

In Figure 1 we also note the appearance of a faint X-ray source, $\sim
7^{\prime}$ due north of \src, which is confined to the hard-band
image. The J2000 coordinates of the serendipitous source in the
registered image are R.A. $ 16^h 17^m 30^s.1$, Dec. $-50{^\circ}
55^{\prime} 05$ degrees. No equivalent source at this location is
apparent in any of the archival X-ray images of \rcw. We checked the
HEASARC catalogs and find no X-ray, radio, optical objects within 
$3^{\prime}$ of the detection. The absence of a source in the soft X-ray
band-passes of the earlier missions may indicate that this source,
which we designate \hbox{AXS~J$161730-505505$}, is highly
absorbed. Alternatively, the lack of a catalog ID suggests the
discovery of a new transient or variable source.

Having demonstrated the presence of \src, we now use \ASCA s full
spectroscopic capability to infer some of its properties.  This is not
straightforward because the spectrum of the point source is
superimposed on strong diffuse emission from the remnant, whose
contribution to the source is increased due to the broad wings of
\asca s PSF. For the GIS, the situation is exacerbated by the reduced
spatial resolution of the GIS detectors. Because of the increased
diffuse emission in the larger extraction region required to sample
the source, all attempts to fit the GIS spectrum produced
unconstrained results, despite the GIS's higher sensitivity at
energies above 2 keV.

Our approach to fitting the SIS spectrum was to first characterize the
remnant spectrum, and to use it as a ``background'' to fit the source
spectrum.  The combined source plus remnant spectrum was extracted from
a small circular region $2.4^{\prime}$ in diameter, centered on the
source to maximize the source signal. We used data from SIS-0
exclusively, as the point source abuts the inter-chip gap on SIS-1.
For SIS-0, most of the source counts were collected on a single CCD
chip.  For a ``background'' field we extracted counts from this
chip using a $1.2^{\prime}$ wide annular segment covering the
bright SNR shell. An average sky background, made up of data from four
surrounding archival blank survey fields, was subtracted from these
spectra.  We also generated custom response files for SIS-0 that take 
into account the restricted region we used. Systematic uncertainties 
are expected at the $\sim 5\%$ level due to the azimuthal energy 
dependence of the PSF.

The shocked nebula plasma is evidently not in ionization equilibrium,
as a single component Raymond-Smith ionization equilibrium plasma
model does not adequately fit the spectrum.  We used instead a
non-equilibrium ionization (NEI) plasma model for Sedov hydrodynamics
assuming $T_e = T_{ion}$ (Hamilton, Sarazin \& Chevalier 1983), in
which the abundances of the elements were allowed to vary with C, N, O
tied to each other and Fe tied to Ni.  The spectrum is well fit with a
single component model with temperature $kT = 0.3$ keV and column
density $\rm{N_H} = 7 \times 10^{21} \ \rm{cm}^{-2}$ (see Fig. 2a and Table
I). The ionization parameter $\eta \equiv n_o^2$ E is $10^{50}
\rm{(cm^{-6}\ ergs)}$, corresponding to an ionization age $n_e t$ of
$6 \times 10^3$ cm$^{-3}$ yr. 
Assuming a distance of 3.3 kpc, Sedov
hydrodynamics gives an age of about 4000 yr for the remnant. This age 
represents an upper limit for a remnant in the adiabatic phase, and is
consistent with the upper end of ages derived by Nugent \etal\ (1984).

We next fit for the source spectrum, using counts extracted from the
region containing emission from the source+remnant.  The thermal
``background'' model alone, with only the normalization allowed to
vary, gives an unacceptable fit (see Fig. 2b and Table I).  As
anticipated from the spectrally resolved images, there is a clear
excess of counts above 2 keV.  We then added in turn to the model one
of three different continuum components --- blackbody, power-law and
thermal bremsstrahlung.  Each of these greatly improves the fit, as
detailed in Table I.  In each case, we fixed the spectral parameters
of the nebula except the Silicon abundance, for which there is
evidence for slight variation with position.  Only the normalization
of the thermal component was allowed to vary.  The column density was
initially held fixed at the fit value for the nebula and then allowed
to vary freely.  All three continuum models yield comparable $\chi^2$.
Except for the blackbody, the fits favor additional $\rm{N_H}$ for the
point source (see Table I). This may be intrinsic absorption to the
source or due to intervening, cold, clumped material near the SNR,
serendipitously in the line of sight. The best fit black body model
gives $\rm{kT} = 0.56$ keV, and the best fit power law model photon
index is $\Gamma =3.2$ for fixed $\rm{N_H}$, compared to $\rm{kT} =
0.57$ keV and $\Gamma =4.6$ for $\rm{N_H}$ freely fit.

The presence of a hard power-law component might signal an underlying
pulsar. We searched the \ASCA\ data for coherent modulation of the
X-ray emission from the source at periods as short as 8 ms by
performing a FFT on the GIS time series extracted from a circular
region of radius 4$'$.  To minimize contamination from the supernova
remnant, only counts with pulse heights corresponding to energies E
$>$ 2.6 keV were used, resulting in a count rate of 0.09 s$^{-1}$, of
which $2/3$ are source counts and the other $1/3$ arise from the
SNR. We used a bin size of 4 ms and split the time series into four
segments because of computational constraints. From the averaged power
spectra of the four segments we find no significant power at 90\%
confidence at any of the frequencies searched. We searched the entire
data set for periods greater than 2 s and find no modulations
greater then 13\% of the mean count rate, the limit of our sensitivity
at the 90\% confidence level. We also folded the light curve at the 69
ms period reported in IAUC 5588 and find no time modulation of the
signal when compared to a null hypothesis ($\chi_{\nu}^2\sim$1.0).

We considered whether the source displays long-term variability by
examining archival data. A total of 13 imaging observations of \rcw\
are available from the \einstein, EXOSAT, and \ROSAT\ missions,
spread-out fairly evenly over the 16 years since the detection of
\src. Of these, three provide a detection: the discovery observation
with the \einstein\ HRI, the \ASCA\ observation presented here, and
the recently released {\it on-axis} \ROSAT\ HRI observation acquired
in Aug 1995. We folded our model fits obtained with the \asca\ spectra
through the respective instrument spectral responses and compared
their count rates using PIMMS (Mukai 1993). We first made comparisons
using the best fit model parameters derived with fixed $\rm{N_H}$ (see
Table I). For the \einstein\ and \rosat\ HRI, whose observations span
16 years, we find that their background subtracted count rates agreed
to within $< 10\%$.  In contrast, all these fits predict a \ASCA\ SIS
count rate substantially higher than that observed, by a factor of
three or more. One reason for the flux discrepancy may be the
possibility that \src\ is more highly absorbed than \rcw. We again
compared count rates this time using the model fit for which $\rm{N_H}$ 
were allowed to vary. The \asca\ models then predict an HRI
count rates $\sim 2$ higher than measured. If we allow $\rm{N_H}$
to vary even higher, it is possible to obtain agreement, suggesting
either strong intrinsic absorption or a factor of $\sim 2$ variability
of the flux.

As discussed in the next section, the relative count rates
are very sensitive to the instrumental energy band-pass, along with the
assumed $\rm{N_H}$. Given the uncertainties in the \ASCA\ spectrum
and flux due to the nebular contamination below 2 keV, it is unclear
that the source varies significantly. A set of internally consistent
measurements (i.e., with the same instrument) will be needed to 
establish variability.

\section{Discussion}

The \ASCA\ observation suggests two reasons why \src\ has been so
difficult to detect in previous observations: its hard spectrum is
obscured in the soft X-ray band by the nebular flux, and intrinsic 
absorption may reduce its soft flux in the ROSAT band-pass.
At energies above $2.5$ keV the source spectrum is clearly dominant
over the soft thermal flux of the nebula (see Fig 2b). Detecting an
unresolved point-source embedded in a significant background is highly
dependent on the beam-size and energy.  The \ROSAT\ band-pass
is limited by the mirror response to $< 2.0$ keV, whereas the
\einstein\ response continues to $ \sim 4.5$ keV. Although the \ASCA\
beam-size is much greater than that of the on-axis HRI PSF (beam-size
$\sim 5^{\prime\prime}$), the increased high energy response up to
$\sim 10$ keV allows direct imaging of the hard component above $ \sim 2$
keV. For the \ROSAT\ HRI observations made with the target
off-axis, the increased beam-size ($\sim 1^{\prime} \ \rm{vs.} \
5^{\prime\prime}$) and mirror vignetting ($\sim 15\%$) reduced the
detectability of the faint \einstein\ HRI source to below threshold
(see Becker 1993). The large intrinsic instrumental blur of the
\einstein\ IPC and the \ROSAT\ PSPC also reduce the point-source
sensitivity.

In the following we reconsider the main observational arguments
concerning a cooling neutron star origin, an accreting binary, or a
plerionic origin for \src. None of these are found to be completely
satisfactory and so we examine alternative origins for the compact
source in \rcw. Based on the \einstein\ discovery observation, \src\
was suggested to be an isolated cooling \NS\ dominated by thermal
emission. Indeed, several SNR contain intriguing cooling \NS\
candidates, including Puppis A (Petre \etal, 1996) and G296.5+10.0
(Mereghetti \etal, 1996).  Although the \ASCA\ spectrum allows a
blackbody solution, the temperature and luminosity are hard to
reconcile with current theoretical models (see \"Ogelman 1995 for a
review), which may well require reconsideration, perhaps by allowing
for an accreted envelope (see Chabrier \etal\ 1997 and references
therein).

The morphology of \rcw\ is inconsistent with a plerion dominated by
synchrotron emission, such as is observed from Crab-like composite
remnants. The central sources in these young ($<1000$ yrs) SNR
are copious radio emitters with highly polarized extended synchrotron 
nebulae driven by a radio pulsar (Dickel \etal\ 1994). Their non-thermal 
cores produce X-ray spectra with hard power-law photon indices of $\sim 2$. 
Alternatively, some young shell-type SNRs are thought to contain a faint 
plerionic core, such as \geleven\ (Vasisht \etal\ 1996). Although the 
remnant of this young SNR is similar to that of \rcw, the embedded radio 
source likely requires a different origin; perhaps the radio source is at a 
different evolutionary stage.

The evidence against an accreting binary is equally compelling. The
luminosity range for a typical accretion-driven low-mass X-ray binary
(LMXB) is $\sim 10^{36 - 38}$ ergs s$^{-1}$, implying a
distance for \src\ far behind those estimated for \rcw. As noted by
Tuohy \etal\ (1983), the soft thermal nebular spectrum is incompatible
with the implied interstellar column density at that distance. For a
binary origin for \src, we would expect to find an optical companion,
perhaps with ultraviolet excess from an accretion disk,
displaying Doppler shifted optical absorption lines. 

The properties of \rcw\ along with its compact source are quite
similar to other radio-quiet compact X-ray sources found near the
center of SNRs probed by \asca. These sources are identified by
their X-ray emission in the hard band-pass, above the shocked thermal
remnant emission.  In particular, \rcw\ is most similar to Kes\sp73.
Both are young, distant Type II Galactic SNRs with comparable thermal
spectra and apparent morphologies. Unlike \geleven, both
\rcw\ and Kes\sp73 contain central radio-quiet X-ray sources with similar
spectral properties (see Gotthelf \& Vasisht 1997). Furthermore, 
neither shows evidence for a radio or optical counterpart, or for 
plerionic X-ray emission. Kes\sp73 contains a 12 s X-ray pulsar which is likely being spun down quickly by virtue of
an anomalously large magnetic field (see Vasisht \& Gotthelf 1997 and
ref. therein). The most striking difference between the two systems is
the luminosity of the compact source.  The younger ($\sim$ 1000 yrs)
and hotter (kT $\sim 0.7$ keV) Kes\sp73 emits a larger fraction of its
flux from the compact source ($\sim 4 \times 10^{35}$ ergs s$^{-1}$)
relative to the shell, as might be expected for a younger system
radiating energy at a higher rate, perhaps due to the large postulated
magnetic field.

The luminosity and age of \src\ are also consistent with an isolated
pulsar radiating gravitational energy. The maximum luminosity for a
$\sim 2000$ yr old, $\sim 10$ s pulsar powered by
spin-down energy is in the range of 
$\sim 10^{33 - 34}$ \hbox{erg s$^{-1}$},
for a $1.4 \rm{M}_\odot$ neutron star.  Perhaps the pulsar has spun
down, radiating residual braking energy and is sub-luminous for a
plerion, but hotter and brighter then a quiescent cooling \NS. If
\rcw\ is typical of SNR with radio quiet, non-plerionic central
compact sources, this may well explain the mysterious absence of
observed natal \NS s associated with Type II SNR. In this scenario
pulsars with strong magnetic fields are spun-down rapidly, losing
their adolescent radio plerionic nebulae and slowly dim. We hope that
further multi-wavelength studies will help elucidate the significance of
this intriguing object.

\begin{acknowledgements}
{\noindent \bf Acknowledgments} --- This work uses data made available
from the HEASARC public archive at GSFC.  We are indebted to J.
Dickel for kindly providing us with his radio maps of \rcw, C. M.
Becker for use of his timing analysis software, and V. M. Kaspi for
discussion. 
\end{acknowledgements}

\clearpage

\begin{deluxetable}{lcccc}
\small
\tablewidth{0pt}
\tablecaption{Fits to the \ASCA\ SIS Spectra \vfill
\label{table I}}
\tablehead{
 \colhead{\hfil Model \hfil} & \colhead{$\chi^2 (\rm{DoF})$} & \colhead{kT or $\Gamma$} & \colhead{$\rm{N_H}$} & \colhead{\hfil Flux \hfil} \nl
  & & & \colhead{($10^{22}$ cm$^{2}$)} & \colhead{\hfil ($\times 10^{-11}$ cgs) \hfil}
}
\startdata
 Nebula NEI  & 77 (77)  & 0.3 & 0.68 & -  \nl
 Source NEI  & 517 (93) &  -  &   -  & -   \nl
\noalign{\vskip 5pt}
 \multispan5{\hfill --- Fits to source with $N_H$ fixed ---\hfill}\nl
\noalign{\vskip 5pt}
 NEI + BB    & 105 (99) & $0.56_{0.53}^{0.59}$ & 0.68 & $0.62$  \nl
 NEI + PL    & 130 (99) & $3.2_{3.0}^{3.4}   $ & 0.68 & $1.8$  \nl
 NEI + BREM  & 110 (99) & $1.6_{1.4}^{1.8}   $ & 0.68 & $1.0$  \nl
\noalign{\vskip 5pt}
 \multispan5{\hfill --- Fits to source with $N_H$ free ---\hfill}\nl
\noalign{\vskip 5pt}
 NEI + BB    & 104 (98) & $0.57_{0.52}^{0.62}$ & $0.5_{0.0}^{1.1}$ & $0.56$ \nl
 NEI + PL    & 106 (98) & $4.6_{4.0}^{5.3}   $ & $3.1_{2.1}^{4.3}$ & $1.7$ \nl
 NEI + BREM  & 103 (98) & $1.2_{1.0}^{1.5}   $ & $1.6_{1.0}^{2.4}$ & $1.6$ \nl
\enddata
\tablenotetext{} {For the nebula fit the abundances of the elements were allowed to be fit with C, N, O tied to each other and Fe tied to Ni. $\rm{Log} \ \eta$ (cm$^{-6}$ ergs)$ = 50$. All subsequent fits used the nebula 
NEI model fixed with only the normalization allowed to vary. $\Gamma$ is the photon spectral index; kT in units of keV. Flux refers to the unabsorbed flux 
of the second component in the $0.5-10$ keV band.}
\end{deluxetable}

\clearpage 

\begin{figure}

\centerline{ {\hfil\hfil
\psfig{figure=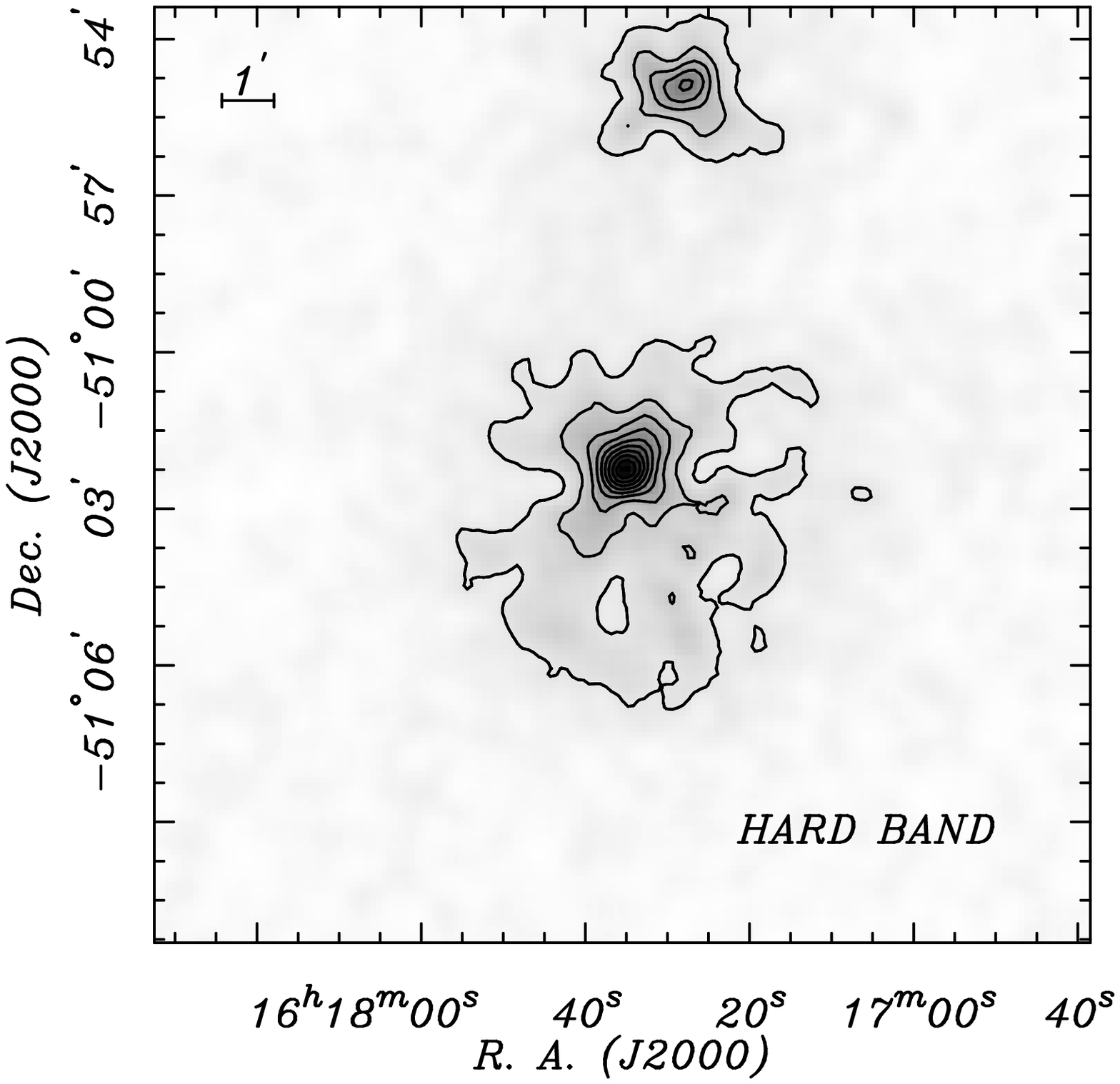,height=4.0truein,angle=270.0,bbllx=25bp,bblly=25bp,bburx=587bp,bbury=520bp,clip=}
\psfig{figure=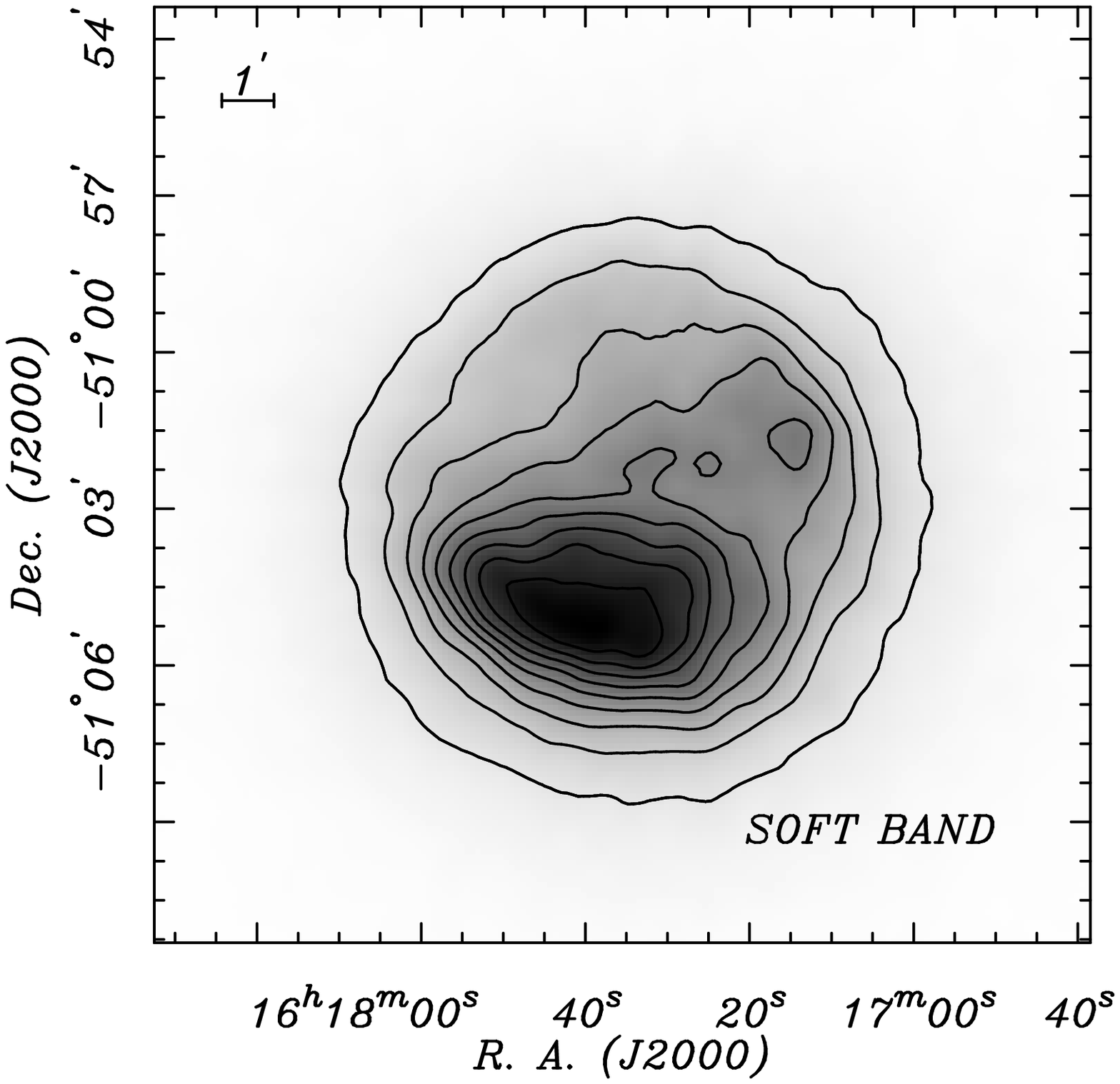,height=4.0truein,angle=270.0,bbllx=25bp,bblly=25bp,bburx=587bp,bbury=520bp,clip=}
\hfil\hfil} }

\caption{Exposure corrected and smoothed \ASCA\ SIS images of \rcw.
a) The hard-band ($> 3.0$ keV) SIS image of \rcw\ shows the 
unambiguous \asca\ detection of \src; the spatial distribution is 
consistent with that of a point source. b) The soft-band ($0.5 - 1.5$ keV) 
image is consistent with previous imaging observations of \rcw. In this 
band-pass the nebula emission dominates the flux from \src. The images 
are displayed with a linear greyscale and contours in 10 equal intervals.}

\end{figure}

\clearpage 

\begin{figure}

\centerline{\psfig{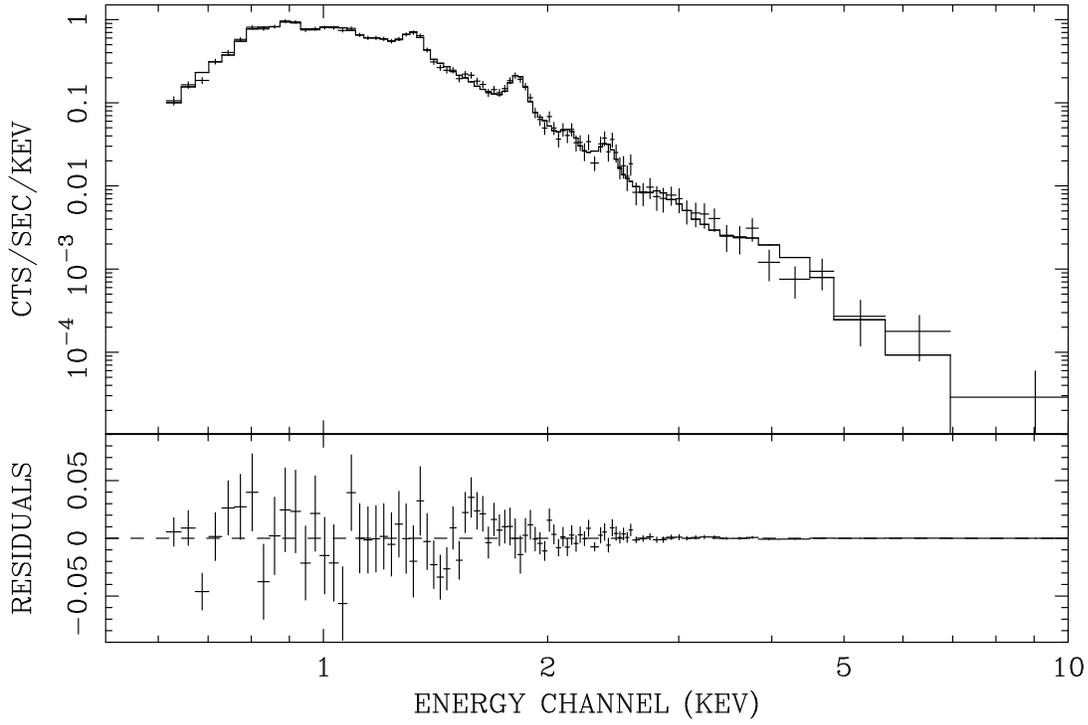}}
\centerline{\psfig{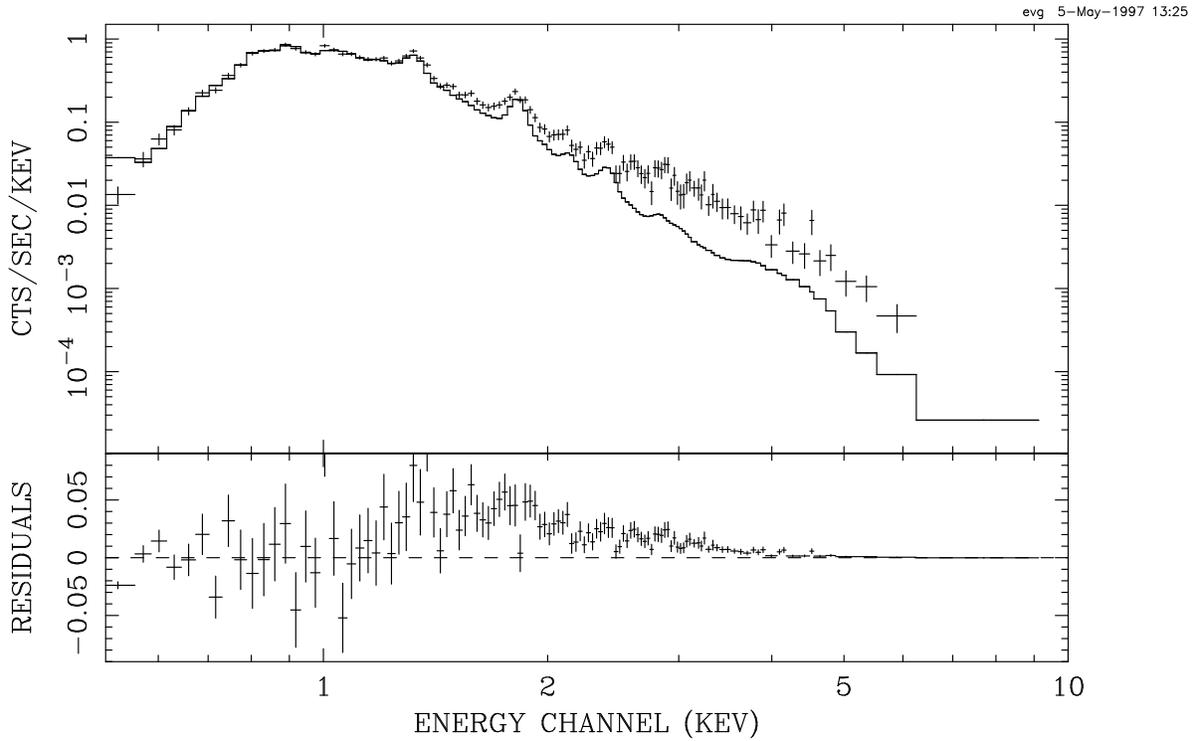}}

\caption{The \asca\ SIS spectrum of \rcw\ showing (a) the 
fitted nebula spectrum and (b) the same model overlaid on the
source spectrum to highlighting the \src\ component.}

\end{figure}

\end{document}